\documentstyle{article}

\title{Quantum control of atomic systems by time-resolved homodyne 
detection of spontaneous emission}

\author{Holger F. Hofmann and Ortwin Hess
        \\Institut f\"ur Technische Physik, DLR,
        \\Pfaffenwaldring 38-40, 70569 Stuttgart, Germany\\[0.2cm]
        G\"unter Mahler 
        \\Institut f\"ur Theoretische Physik und Synergetik,
        \\ Pfaffenwaldring 57, 70550 Stuttgart, Germany}

\begin{document}

\maketitle

\begin{abstract}
We describe the light-matter interaction of a single two level atom with the
electromagnetic vacuum in terms of field and dipole variables by considering
homodyne detection of the emitted fields. Spontaneous emission is then 
observed as a continuous fluctuating force acting on the atomic dipole.
The effect of this force may be compensated and even reversed by feedback.
\end{abstract}

\section{Introduction}
The spontaneous emission of light from a single atom is usually described 
as the random appearence of a photon in the vacuum light field. However, 
this description is only valid if photons are actually detected. 
The field variables continuously evolve from the dipole dynamics of the
atom according to Maxwells equations. If field variables are measured, 
the spontaneous emission of a single two level atom may be interpreted as 
the interaction of a  fluctuating dipole with a noisy light field
\cite{Hof98a,Hof98b}.
Quantum jumps are avoided and the continuous evolution of the atomic system 
may be controlled by weak coherent feedback
fields compensating the observed quantum noise. 
In the following we describe
the evolution of the quantum state of a two level atom conditioned by
projective homodyne detection and discuss some of the possible feedback 
scenarios. 

\section{Homodyne detection of weak fields}

In a balanced homodyne detection setup the coherent laser field of a local
oscillator interferes with the low intensity input field at a beamsplitter
as shown in Figure 1.
\begin{figure}
\begin{picture}(300,210)
\put(135,75){\line(1,1){30}}
\put(165,117){\makebox(50,20){beam}}
\put(165,105){\makebox(50,20){splitter}}

\put(141,90){\line(-1,1){10}}
\put(141,90){\line(-1,-1){10}}
\put(90,90){\line(1,0){51}}
\put(78,90){\circle{24}}
\put(55,55){\makebox(50,20){source}}

\put(220,90){\line(-1,1){10}}
\put(220,90){\line(-1,-1){10}}
\put(159,89){\line(1,0){60}}
\put(159,91){\line(1,0){60}}
\put(219,102){\line(1,0){12}}
\put(219,78){\line(1,0){12}}
\put(231,90){\line(1,0){12}}
\put(231,78){\line(0,1){24}}
\put(210,55){\makebox(50,20){detector 2}}

\put(150,81){\line(1,-1){10}}
\put(150,81){\line(-1,-1){10}}
\put(148,78){\line(0,-1){42}}
\put(152,78){\line(0,-1){42}}
\put(141,7){\framebox(18,29)}
\put(168,22){\makebox(50,20){local}}
\put(168,10){\makebox(50,20){oscillator}}

\put(150,160){\line(1,-1){10}}
\put(150,160){\line(-1,-1){10}}
\put(149,99){\line(0,1){60}}
\put(151,99){\line(0,1){60}}
\put(138,159){\line(0,1){12}}
\put(162,159){\line(0,1){12}}
\put(150,171){\line(0,1){12}}
\put(138,171){\line(1,0){24}}
\put(168,173){\makebox(50,20){detector 1}}
\end{picture}
\caption{Schematic setup of balanced homodyne detection}
\end{figure}
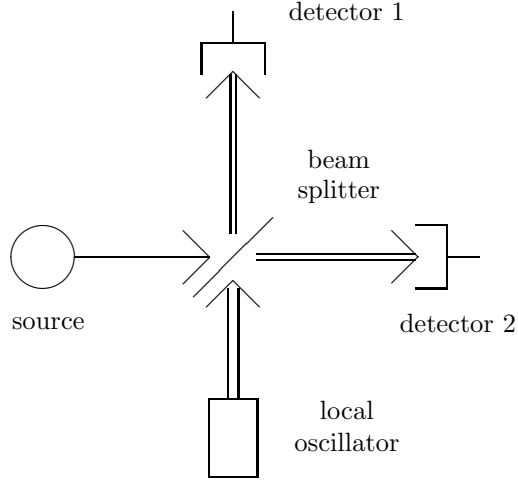 
The difference between the photon numbers registered in detector 1 and 
detector 2 corresponds to the interference term
\begin{equation}
\Delta\hat{n} = \hat{a}^\dagger\hat{b} + \hat{a}\hat{b}^\dagger.
\end{equation}
where $\hat{a}$ and $\hat{b}$ represent the annihilation operators of
the local oscillator mode and the source field, respectively. The
field modes are emitted during the measurement time interval $\tau$
and represent wave packets of length $c\tau$.
The quantum state emitted by the local oscillator may be represented by
a coherent state with an average complex amplitude of
$\alpha$. The photon number difference $\Delta n$ then corresponds to
\mbox{$2\mid\alpha\mid$} times the quadrature component of the source field
which is in phase with the local oscillator. For weak fields, the 
probability distribution of the measurement results $\Delta n$ is
approximately given by the vacuum fluctuations of the observed quadrature
component, 
\begin{equation}
p(\Delta n) \approx \frac{\exp[-\frac{\Delta n^2}{2\mid \alpha^2 \mid}]}
{\sqrt{2\pi\!\mid\!\alpha^2\!\mid}}.
\end{equation}
Within the measurement time interval $\tau$ the dipole fluctuations of a 
two level atom with a spontaneous emission rate of $\Gamma$ emit an 
average light field energy of $\Gamma\tau$ times the quantum fluctuation
intensity of $\hbar\omega/2$. If $\Gamma\tau$ is much smaller than one,
the dipole radiation emitted by the atom is much weaker than the quantum
fluctuations of the electromagnetic vacuum. Therefore, the dipole radiation
is obscured by quantum noise and the information about the state of the
atomic system obtained in the homodyne detection measurement is extremely
small. Nevertheless some information is obtained about the most likely
orientation of the atomic dipole and this observation will modify the 
quantum state of the system as explained below.  

\section{Quantum diffusion of a two level atom}

The back action of continuous time-resolved homodyne detection on a 
quan\-tum system results in a stochastic evolution of the wave function
equivalent to a Monte Carlo wavefunction formalism \cite{Wis93,Car96}.
The derivation from a projective measurement base is discussed
in \cite{Hof98a}. It is convenient to describe the back action in 
terms of the Bloch vector ${\bf s}$ of the atomic two level system,
where $s_x$ is the expectation value of the observed dipole component 
and $s_z$ is the expectation value of the atomic inversion.
The $s_y$ component describes the expectation value of the
unobserved dipole component. 
The back action corresponding to a measurement result of $\Delta n$
for an arbitrary initial Bloch vector $(s_x,0,s_z)$ in the $s_y=0$ plane
reads 
 
\begin{samepage}
\begin{eqnarray}
\delta s_x =& &\sqrt{\Gamma\tau}\frac{\Delta n}{\mid\alpha\mid}
               (1 + s_z) s_z \nonumber \\
\delta s_z =&-&\sqrt{\Gamma\tau}\frac{\Delta n}{\mid\alpha\mid}
               (1 + s_z) s_x.
\end{eqnarray}
\end{samepage}
The Bloch vector is thus rotated by an angle of
$\sqrt{\Gamma\tau}\frac{\Delta n}{\mid\alpha\mid}(1+s_z)$ around the
y-axis in response to the homodyne detection measurement.  
\section{Controlling the quantum state by feedback}

Without feedback, the ground state $s_z=-1$ is stationary while diffusion is
at a maximum for the excited state $s_z=+1$. It is possible to interpret
this back action effect as a sum of Rabi rotations induced by the quantum
noise and an epistemological effect of the information obtained about the
dipole component $s_x$ of the atom. The ground state is stationary 
because the dipole emission effects compensate
the absorption of vacuum fluctuations. In the excited state the dipole
emission effect and the response to the vacuum fluctuations add up and
cause twice the diffusion expected from classical field fluctuations.

By applying a negative
feedback equal to the observed quadrature component the Rabi rotations
induced by the quantum fluctuations
of the measured field component may be compensated. The diffusive back action
which remains is then associated with the information gained about
the atomic system. As mentioned in the previous section,
it is possible to
identify this back action as a weak measurement effect of the dipole variable
$s_x$. Due to the weak dipole radiation emitted by the atom, $\Delta n>0$ is 
more likely for $s_x=+1$ and $\Delta n<0$ is more likely for $s_x=-1$. 
Coherent superpositions of dipole eigenstates diffuse because of the modified
statistical weight of the dipole eigenstate components. 
Consequently the dipole eigenstates are 
stationary while the diffusion is at a maximum for both the ground state
$s_z=-1$ and the excited state $s_z=+1$. 

By applying twice the negative feedback necessary for compensation it
is possible to invert the effects of quantum fluctuations. The excited state
$s_z=+1$ becomes stationary and diffusion is at a maximum in the ground 
state $s_z=-1$. By inverting the sign of the observed quantum fluctuation
component the roles of the excited state and the ground state are exchanged. 
The excited state now absorbs quantum fluctuations, thus compensating
the effects of dipole emission, while the ground state amplifies the 
fluctuations and thus shows twice the average diffusion.

The back action effect of homodyne detection without feedback, with
feedback compensating the quantum fluctuations and with feedback inverting
the quantum fluctuations is shown in figure 2.
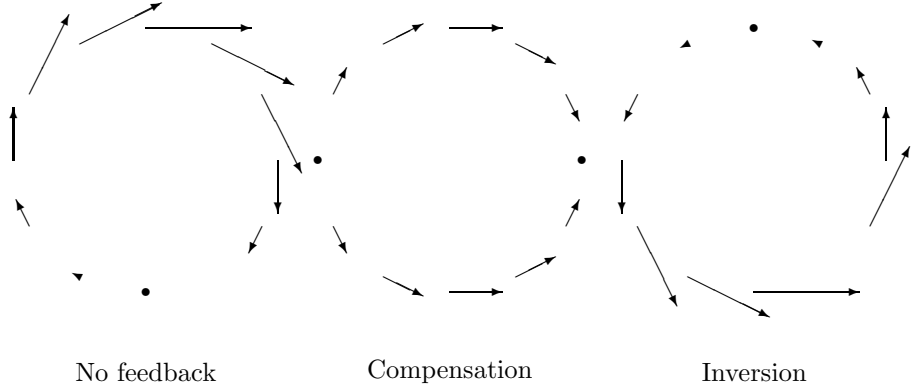
\begin{figure}
\begin{picture}(350,180)
\put(25,10){\makebox(50,20){No feedback}}
\put(50,50){\circle*{3}}
\put(25,56){\vector(-2,1){3}}
\put(6,75){\vector(-1,2){5}}
\put(0,100){\vector(0,1){20}}
\put(6,125){\vector(1,2){15}}
\put(25,144){\vector(2,1){31}}
\put(50,150){\vector(1,0){40}}
\put(75,144){\vector(2,-1){31}}
\put(94,125){\vector(1,-2){15}}
\put(100,100){\vector(0,-1){20}}
\put(94,75){\vector(-1,-2){5}}

\put(140,10){\makebox(50,20){Compensation}}
\put(165,50){\vector(1,0){20}}
\put(140,56){\vector(2,-1){15}}
\put(121,75){\vector(1,-2){5}}
\put(115,100){\circle*{3}}
\put(121,125){\vector(1,2){5}}
\put(140,144){\vector(2,1){15}}
\put(165,150){\vector(1,0){20}}
\put(190,144){\vector(2,-1){15}}
\put(209,125){\vector(1,-2){5}}
\put(215,100){\circle*{3}}
\put(209,75){\vector(1,2){5}}
\put(190,56){\vector(2,1){15}}

\put(255,10){\makebox(50,20){Inversion}}
\put(280,50){\vector(1,0){40}}
\put(255,56){\vector(2,-1){31}}
\put(236,75){\vector(1,-2){15}}
\put(230,100){\vector(0,-1){20}}
\put(236,125){\vector(-1,-2){5}}
\put(255,144){\vector(-2,-1){3}}
\put(280,150){\circle*{3}}
\put(305,144){\vector(-2,1){3}}
\put(324,125){\vector(-1,2){5}}
\put(330,100){\vector(0,1){20}}
\put(324,75){\vector(1,2){15}}
\end{picture}

\caption{Diffusion of the Bloch vector corresponding to measurement results
of $\Delta n>0$ for different feedback scenarios. The top of the circles 
represents the excited state $s_z=+1$ and the bottom represents the ground 
state $s_z=-1$.}
\end{figure}

\section{Conclusions}

The coherent and excited states of a two level atom may be stabilized by
homodyne detection and negative feedback. If the dipole
eigenstates are stabilized the back action of homodyne detection corresponds 
to weak measurements of the dipole component which emits light in phase with 
the local oscillator. The irreversible nature of spontaneous emission
is thus associated with a weak projective measurement of the atomic 
dipole. 

Homodyne detection avoids the discontinuous quantum jumps associated with
photon detection. Therefore the stabilization of quantum states does not
require short time pulses of high intensity. The feedback amplitudes 
needed for the stabilization of quantum states by homodyne detection
and feedback are of the same 
order of magnitude as the observed vacuum fluctuations. It may thus be
sufficient to couple the local oscillator field to the atomic system 
as a function of the intensity difference $\Delta n$ by an optical
nonlinearity.

\end{document}